\documentclass[pdflatex,sn-mathphys-num]{sn-jnl}

\usepackage{graphicx}%
\usepackage{multirow}%
\usepackage{amsmath,amssymb,amsfonts}%
\usepackage{amsthm}%
\usepackage{mathrsfs}%
\usepackage[title]{appendix}%
\usepackage{xcolor}%
\usepackage{textcomp}%
\usepackage{manyfoot}%
\usepackage{booktabs}%
\usepackage{algorithm}%
\usepackage{algorithmicx}%
\usepackage{algpseudocode}%
\usepackage{listings}%

\theoremstyle{thmstyleone}%
\theoremstyle{thmstyletwo}%

\theoremstyle{thmstylethree}%

\begin{document}

\title[Article Title]{Visualizing shear-induced structures in carbon black gels by tomo-rheoscopy}


\author*[1,2,3]{\fnm{Julien} \sur{Bauland}}\email{julien.bauland@univ-lyon1.fr}

\author[1]{\fnm{Stéphane G.} \sur{Roux}}\email{stephane.roux@ens-lyon.fr}

\author[4]{\fnm{Stefan} \sur{Gst{\"o}hl}}
\email{stefan.gstoehl@hest.ethz.ch}

\author[4]{\fnm{Christian M.} \sur{Schlep{\"u}tz}}\email{christian.schlepuetz@psi.ch}

\author[5]{\fnm{Michael} \sur{Haist}} \email{michael.haist@baustoff.uni-hannover.de}

\author*[1,6]{\fnm{Thibaut} \sur{Divoux}}\email{thibaut.divoux@ens-lyon.fr}

\affil[1]{\orgname{ENSL, CNRS, Laboratoire de physique}, \orgaddress{F-69342 Lyon, France}}

\affil[2]{Department of Materials, ETH Zürich, Zürich, Switzerland}

\affil[3]{Université Claude Bernard Lyon 1, CNRS, Institut Lumière Matière, UMR5306, Villeurbanne F69100, France}

\affil[4]{Swiss Light Source, Paul Scherrer Institut, Villigen, Switzerland}


\affil[5]{Leibniz Universität Hannover, Hannover, Germany}

\affil[6]{International Research Laboratory, French American Center for Theoretical Science, CNRS, KITP, Santa Barbara, 93106-4030 USA}


\abstract{Suspensions of attractive particles form space-spanning networks that endow the suspension with solid-like behavior at rest. The microstructure of these colloidal gels depends sensitively on the shear history and on the path followed across the sol-gel transition, resulting in viscoelastic properties that can be tuned by shear. Here, we report \textit{in situ} X-ray tomo-rheoscopy experiments on carbon black gels whose elastic properties exhibit a non-monotonic dependence on the shear intensity applied prior to flow cessation. By directly imaging the gel microstructure under a well-controlled rheological protocol, we reveal the emergence of pronounced structural heterogeneities extending from tens to hundreds of microns -- length scales far larger than those accessible by conventional scattering techniques such as Ultra-Small Angle X-ray Scattering. In particular, we show that only the low-shear reinforcement of elasticity correlates with a growing mesoscale correlation length, while high-shear strengthening occurs without detectable mesoscale reorganization.
These observations demonstrate that flow memory in colloidal gels is not solely governed by local particle rearrangements, but is also encoded in a mesoscale structural organization extending up to 100 times the particle size. More broadly, this work highlights the power of X-ray tomo-rheoscopy to uncover large-scale structural signatures of flow history in soft materials, opening new perspectives to tailor their mechanical properties.}


\maketitle

\section{Introduction}\label{sec:intro}

Disordered soft solids -- from colloidal gels to soft glasses such as dense emulsions -- exhibit a wide variety of mechanical behaviors arising from the interplay between structural disorder and slow dynamics across multiple lengths and time scales \cite{Nicolas:2018}. Even at rest, these materials evolve spontaneously through slow out-of-equilibrium relaxation processes, commonly referred to as physical aging, reflecting a gradual relaxation toward deeper metastable states \cite{Joshi:2014,Joshi:2018,Jain:2020,Bauland:2024b}. 
Under external forcing, aging dynamics can become \textit{directed}, as the applied stress biases the exploration of configuration space and promotes anisotropic structural evolution \cite{Pashine:2019,Hexner:2020}.
At larger deformations, disordered soft solids display nonlinear responses culminating in yielding and flow, accompanied by pronounced structural reorganization \cite{Divoux:2012,Bonn:2017,Divoux:2024}.  

Among soft solids, colloidal gels stand out as materials whose solid-like properties originate from a percolated network of attractive constituents \cite{Trappe:2004,Zaccarelli:2007}. Remarkably, such networks form at very low particle concentrations (typically 0.1--10\% w/w). In contrast to colloidal glasses, where dense packing constrains structural configuration, colloidal gels maintain an open and heterogeneous microstructure \cite{Dibble:2006,Johnson:2019,Hsiao:2014}. This microstructure, together with interparticle attractions comparable to or exceeding thermal energy, gives rise to a large number of metastable configurations, between which the system can be driven under external stimuli. Consequently, the structure and rheology of colloidal gels can be controlled by external stimuli, such as shear \cite{Koumakis:2015,Dages:2022,Das:2022,Nelson:2022,Sudreau:2022b,Burger:2024,Colombo:2025}, ultrasound~\cite{Pandey:2017,Dages:2021,SaintMichel:2022,Athanassiadis:2022}, electromagnetic fields \cite{Tasoglu:2014,Munteanu:2025}, or light \cite{Jones:2016}, making them appealing model systems and promising candidates for the design of responsive ``smart'' materials.  

Under shear, the coupling between flow and microstructure makes colloidal gels highly sensitive to flow history and prone to so-called memory effects, whereby information about past shear is encoded in the microstructure and manifested through the associated residual stresses \cite{Fiocco:2014,Moghimi:2017b,Schwen:2020}.
Experimental studies exploring this rheological memory have shown that shear-induced structuring can extend  well beyond the particle scale in systems as diverse as PMMA particles, boehmites, silica particles dispersed in water, and carbon black dispersed in oil~\cite{Sudreau:2023,Bauland:2025,Jiang:2022,Colombo:2025}. These studies point to the emergence of structural organization over length scales much larger than the size of individual particles, although the range of experimentally accessible length scales remains strongly dependent on the probing technique.
This pronounced sensitivity to shear history makes establishing structure--property relationships particularly challenging, as it requires probing length scales ranging from nanometers to millimeters. Additionally, colloidal gels are prone to flow heterogeneities, such as fracture or shear banding, leading to spatial variations of the microstructure over length scales comparable to the gap size \cite{Varadan:2003,Perge:2014,Grenard:2014,Bouzid:2020}.  

At nanometer to micrometer scales, indirect methods such as Small-Angle Scattering using X-rays~\cite{Dages:2022} or neutrons~\cite{Hipp:2019,Richards:2019}, have proven highly effective to investigate yielding and shear-induced structures when coupled to a rheometer, while X-ray photon correlation spectroscopy (XPCS) provides complementary access to the slow dynamics of these systems \cite{Leheny:2015}. In practice, synchrotron-based scattering experiments typically access length scales from a few nanometers to a few microns, the upper limit being set by the sample-detector distance. At larger scales, from microns to millimeters, mesoscopic structures in soft solids have been investigated using small-angle light scattering~\cite{Aime:2018}, wide-field optical techniques such as light-sheet fluorescence microscopy~\cite{Spicer:2025}, and confocal microscopy \cite{Colombo:2019}. However, these approaches require optically transparent samples, which severely restricts the range of materials that can be studied.  

X-ray computed tomography (CT) overcomes this limitation by taking advantage of the high penetration depth of X-rays compared to visible or UV light, enabling three-dimensional imaging of optically turbid samples~\cite{Maire:2014,Withers:2021}. Contrast in X-ray CT can arise either from differences in electronic density (absorption contrast), or from refractive index variations (phase contrast). The latter requires coherent X-ray beams available at synchrotron sources and is particularly well suited for soft materials with weak absorption contrast, such as biological tissues. In practice, tomographic reconstructions are obtained from a series of two-dimensional projections acquired while rotating the sample. With typical voxel sizes ranging from 0.2--10~$\mu$m, X-ray CT is ideally suited to probe intermediate length scales that remain difficult to access by other techniques.  

Beyond early custom-built devices~\cite{Deboeuf:2018}, the development of rheometers equipped with dual air-bearing motors has enabled tomo-rheoscopy experiments that combine \emph{in situ} rheological measurements with three-dimensional X-ray imaging under controlled shear histories. A controlled differential rotation between the upper and lower tools imposes shear while the sample undergoes a net solid-body rotation, thereby enabling simultaneous rheological measurements and tomographic imaging. Time-resolved tomo-rheoscopy studies have captured, for instance, stress buildup and redistribution in foams, as well as the flow behavior of cement pastes during shear start-up \cite{Schott:2025,Link:2023}.  

In this article, we combine X-ray tomo-rheoscopy with controlled shear protocols to directly visualize how flow history imprints the microstructure of carbon black gels at \textit{mesoscopic} length scales. We show that the non-monotonic dependence of the gel elastic modulus on the pre-shear rate --recently reported in \cite{Bauland:2026}-- is systematically associated with the emergence of large-scale structural heterogeneities, extending from a few tens to several hundreds of micrometers. These heterogeneities exhibit pronounced spatial gradients across the shear cell and, depending on particle concentration and shear history, may also display complex three-dimensional organization. Our results demonstrate that shear-induced memory in colloidal gels is not encoded solely at the scale of individual clusters or interparticle bonds, but instead manifests through a mesoscale structural organization that bridges local aggregation mechanisms and macroscopic mechanical response.

The remainder of the article is organized as follows. In Section~\ref{sec:MM}, we describe the carbon black dispersions, the X-ray tomo-rheoscopy setup, and the image analysis workflow. Section~\ref{sec:results} presents the combined rheological and structural results for two particle concentrations, highlighting the emergence of shear-dependent structural heterogeneities and their spatial organization. Finally, Section~\ref{sec:discussion} discusses the implications of these findings for flow-induced memory effects in colloidal gels and outlines perspectives for extending tomo-rheoscopy to other soft glassy materials.

\section{Materials and Methods}
\label{sec:MM}

\subsection{Carbon Black dispersions}

\begin{figure*}[t!]
\centering
\includegraphics[scale=0.5, clip=true, trim=0mm 20mm 0mm 20mm] {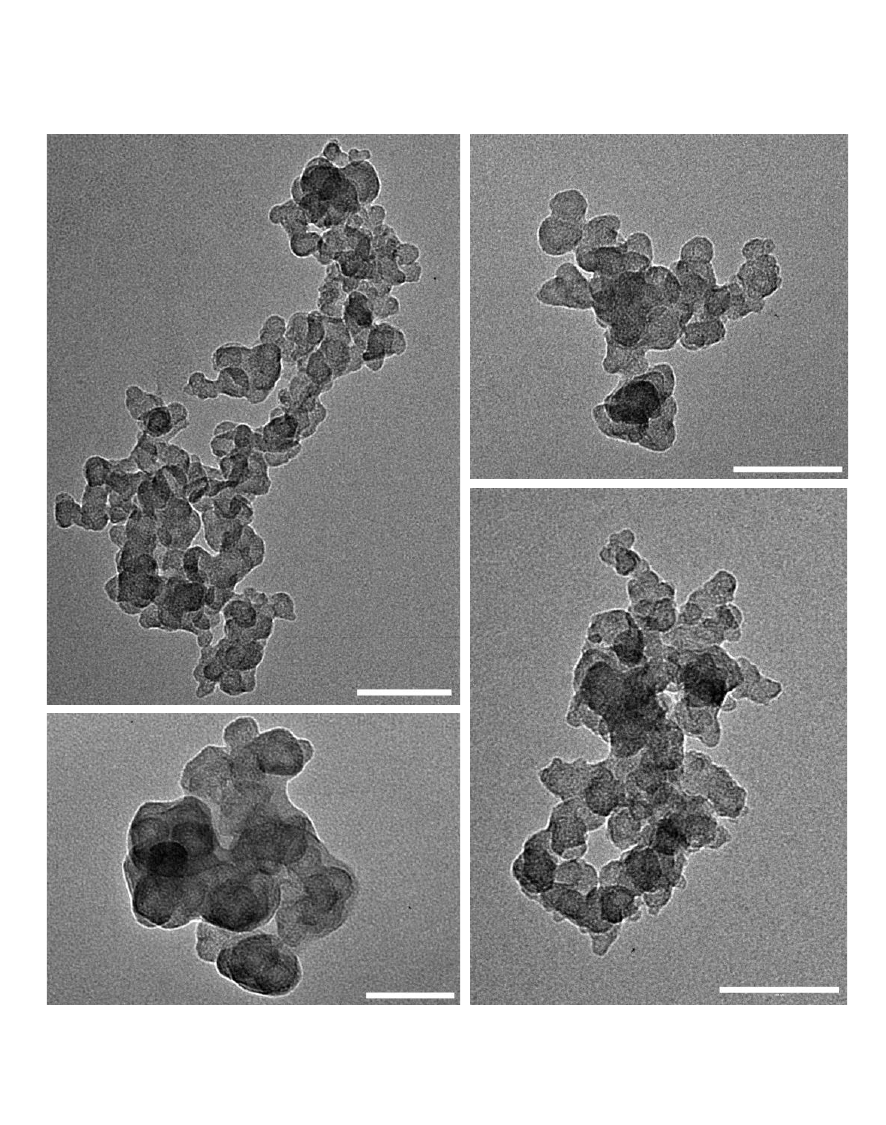}
\caption{Representative Transmission Electron Microscopy (TEM) images of individual carbon black (Vulcan PF) particles. Scale bar is $100~\rm{nm}$. Samples were prepared by depositing a 5 \textmu L drop of a dilute ethanol dispersion of carbon black onto a carbon film (EMS CF300-Cu-UL Carbon Support Film), followed by drying in a dust-free environment.}
\label{fig:TEM}
\end{figure*}

Carbon Black (CB) particles used in this study are Vulcan\textsuperscript{\textregistered} PF (Cabot), with a material density $d_{\rm CB} = 2.26 \pm 0.03$. As shown in Figure~\ref{fig:TEM}, where transmission electron micrographs (JEOL JEM-1400 TEM) of individual particles are displayed, this specific CB grade consists of polydisperse, fractal, and unbreakable particles with a characteristic radius of approximately $100~\rm nm$ and a fractal dimension $d_f = 2.85$. These particles are themselves composed of primary nodules with an average radius of $20~\rm nm$ \cite{Bauland:2024}. 

Following established dispersion protocols \cite{Dages:2022,Bauland:2024,Bauland:2025}, CB particles were dispersed in mineral oil (RTM17 Mineral Oil Rotational Viscometer Standard, Paragon Scientific) with viscosity $\eta_f = 252.1~\mathrm{mPa.s}$ at $T = 25^{\circ}$C, and density $d_{\rm oil} = 0.871$. After hand mixing, the samples were sonicated during 2~h in an ultrasonic bath (Ultrasonic cleaner, DK Sonic\textsuperscript{\textregistered}, United-Kingdom) to ensure complete dispersion. Two mass fractions, $c_w = 3$ and $6~\%$ ($w/w$), were investigated, corresponding to CB volume fractions $\phi = 1.2$ and $2.4~\%$ ($v/v$), respectively. 

In apolar solvents such as mineral oil, carbon black particles interact through short-range attractive forces. As a result, they readily form space-spanning gel networks at low volume fractions, making these systems well-suited model colloidal gels for investigating the interplay between flow, structure, and mechanical properties \cite{Youssry:2013,Ovarlez:2013,Helal:2016,Varga:2019,Richards:2023,Dages:2022}.

\subsection{X-ray tomo-rheoscopy}
\label{sec:Xraytomo}

\subsubsection{Synchrotron imaging and tomo-rheoscopy set-up}
X-ray tomo-rheoscopy experiments were conducted at the TOMCAT Beamline X02DA of the Swiss Light Source (SLS), Paul Scherrer Institute (Villigen, Switzerland) \cite{Stampanoni:2006}. The X-ray beam, produced by a 2.9~T superbending magnet, was monochromatized to an energy of $12.0$~keV using a Ru/C double multilayer monochromator. The experimental end station was located approximately $25~\rm m$ from the source. 

Radiographic imaging was performed using a high-numerical-aperture microscope with 4x magnification \cite{Buhrer:2019}, coupled with a $150~\mathrm{\mu m}$ thick LuAG:Ce scintillator positioned about $400~\rm mm$ downstream of the sample. The magnified visible-light image generated by the scintillator was recorded using the GigaFRoST camera system \cite{Mokso:2017}, featuring 2016 x 2016 pixels with a native pixel size of $11~\mathrm{\mu m}$. This configuration resulted in an effective pixel size of $2.75~\mathrm{\mu m}$ for the radiographic images, corresponding to a horizontal field of view of $5.544~\rm mm$. In the vertical direction,  images were cropped to 1228 pixels, corresponding to a field of view of $3.377~\rm mm$, as the synchrotron X-ray beam intensity dropped off significantly outside this region.

\begin{figure*}[t!]
\centering
\includegraphics[width=0.5\columnwidth]{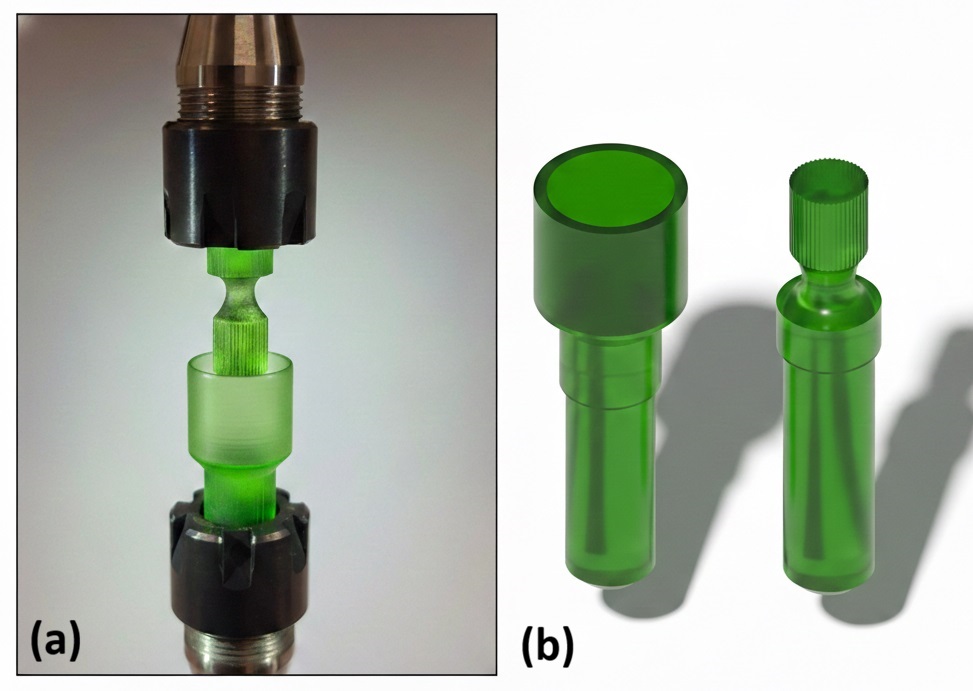}
\caption{(a) Photo of the custom-made Couette cell geometry mounted on the dual motor rheometer. (b) Artificial rendering of the CAD design for the cup and bob, pictured separately. The scale is set by the outer cylinder inner diameter of $R_o=10~\rm mm$.}
\label{fig:Setup}
\end{figure*}

The samples were sheared using a dual-motor rheometer (Anton Paar MCR 702e TwinDrive\textsuperscript{\textregistered}) equipped with a small custom-built, X-ray transparent Couette geometry allowing full tomographic imaging of the gap volume. The geometry consisted of coaxial cylinders of inner and outer radii $R_i = 3~\mathrm{mm}$ and $R_o = 5~\mathrm{mm}$, respectively, corresponding to a gap width of $2~\rm mm$. The outer cylinder (cup) had a depth of $10~\rm mm$, while the inner cylinder (bob) featured an active rheological height of $6.0~\rm mm$ and was serrated with asperities of $100~\mathrm{\mu m}$ to minimize wall slip. 

The geometry was fabricated via stereolithography using a Photon Mono SE 3D printer (Anycubic) equipped with a 405~nm LCD-masked UV source and a lateral (XY) resolution of 50~$\mu$m. A standard translucent green photopolymer resin (diacrylate/methacrylate system, Anycubic) was used. After manufacture, the parts were thoroughly rinsed in isopropanol and post-cured under UV light for 6~min to ensure complete polymerization. 

Both cylinders were mounted on $7~\rm mm$ diameter shafts, and aligned on the rheometer rotation axes using the SCF7 multi-purpose torsional fixture, as shown in Fig.~\ref{fig:Setup}. Prior to measurements, the cup was partially filled with the carbon black dispersion, and the inner cylinder was lowered into position, leaving a gap distance of $1~\rm mm$ to the bottom of the cup, as determined from zero-gap calibrations. Then, the carbon black dispersion was topped up until its surface level reached the upper edge of the serrated bob to achieve well-defined rheological measurements.

Tomographic imaging of the samples at rest was performed by imposing a co-rotation of the inner and outer cylinders at an angular speed of $9^{\circ}$/s, such that the sample underwent a rigid body rotation around the rheometer axis without experiencing any additional shear deformation. To image the full horizontal extent of the sample, extended field of view scans were employed, by positioning the projected rotation axis near the left edge of the camera image and rotating the sample over $360^\circ$. For each scan, 4000 radiographic projections were acquired at a frame rate of $100~\rm Hz$ (exposure period $10~\rm ms$, exposure time $9.995~\rm ms$ to account for the readout time of the imaging chip), resulting in a total scanning time of $40~\rm s$ per tomographic volume. During tomographic acquisition, the sample microstructure was assumed to be stationary on the timescale of the scan, i.e., structural evolution due to aging or relaxation was negligible over the 40 s acquisition time.

\subsubsection{Tomographic reconstruction and spatial resolution}
Tomographic reconstruction was performed using a non-iterative algorithm (gridrec)~\cite{Marone:2012}, combined with single-distance phase retrieval based on the Paganin approach \cite{Paganin:2002}, using a ratio between the real and imaginary parts of the refractive index decrement of $\delta/\beta = 100$, which was chosen by visual inspection to optimize the gray level contrast. Reconstructions yielded 1228 slices of 2016 x 2016 voxels with an isotropic voxel size of $2.75~\mathrm{\mu m}$. This spatial resolution enables direct visualization of structural heterogeneities over length scales from a few micrometers up to several hundreds of micrometers, well beyond the upper length scale accessible to ultra-small-angle X-ray scattering, which typically probes structures up to a few microns \cite{Narayanan:2022}.

\subsubsection{Shear protocol and rheological measurements}
The rheometer was operated in TwinDrive\textsuperscript{\textregistered}) mode using a counter-movement configuration, where both the upper and lower motors rotated simultaneously in opposite directions at a 50:50 motion ratio. Prior to imaging, CB dispersions were rejuvenated by applying a shear rate $\dot{\gamma} = 300~\mathrm{s}^{-1}$ for $60~\mathrm{s}$. Samples were then subjected to a constant pre-shear rate $\dot{\gamma}_0$ for $200~\mathrm{s}$. After flow cessation, the time evolution of the CB suspension linear viscoelastic properties was monitored over $300~\mathrm{s}$ using small-amplitude oscillatory shear at a strain amplitude $\gamma = 0.1~\%$ and an angular frequency $\omega = 2\pi~\mathrm{rad \cdot s^{-1}}$. Tomographic scans were subsequently performed on the quiescent samples. This protocol was repeated for various pre-shear rates $\dot \gamma_0$.

\subsection{Image processing workflow and correlation analysis}
\label{sec:imageanalysis}

Each reconstructed image $I(x,y)$ is processed using the following workflow. First, the radial intensity profile is determined by converting the image to polar coordinates $I(r,\theta)$ and averaging over the angular coordinate, $m(r) = \frac{1}{N_\theta}\sum_\theta I(r,\theta)$. This radial profile is then mapped back to Cartesian coordinates to construct a radial mean image $m(x,y)$ of the same size as the original image. The radial mean is subsequently subtracted from the raw image to obtain a corrected image with zero radial mean, $\tilde{I}=I(x,y) - m(x,y)$. To suppress bias from high-intensity features associated with impurities visible on the images and originally present in the carbon black powder, pixels with intensities larger than five times the standard deviation of $\tilde I$ are removed. These outliers represent less than 1~\% of the total image area (see Fig.~\ref{figsupp:outliers} in Appendix~\ref{sec:outliers}).

For quantitative analysis, the processed image is then divided into four regions of interest corresponding to the whole gap, as well as the inner, middle, and outer portions of the gap, whose radial extents are defined in Section~\ref{sec:1p2CB}. For each region, we compute a radial two-point correlation function
\begin{equation}
\rho(r,\theta) = \frac{1}{N_a} \sum_a \frac{\tilde{I}(a, \theta) \tilde{I}(a+r,\theta)}{\mbox{Var}_{\tilde{I}}}
\end{equation}
where the sum runs over all radial positions $a$, and $\mbox{Var}_{\tilde{I}}$ corresponds to the variance of $\tilde I$ over the region of interest. Because the correlation function varies primarily along the radial direction and shows no significant dependence on the angular coordinate $\theta$, it is further averaged over all angles to obtain the radial correlation coefficient
\begin{equation}
    \rho(r) =  \frac{1}{N_\theta} \sum_\theta \rho(r,\theta)
\end{equation}
This analysis provides a quantitative measure of the characteristic structural length scales within the reconstructed volumes, which will be used in Section~\ref{sec:results} to characterize the evolution of the gel microstructure under different pre-shear conditions. 

For visual display, outliers are conserved. Images are colored by normalizing intensities to the 99.5~th percentile, mapping the data into the range [0,1] to clip the brightest pixels while preserving structural contrast. The resulting values are then mapped to the ``magma'' colormap from the ``200 colormaps'' collection~\cite{Liu2026}.

\begin{figure}[t!]
\centering
\includegraphics[scale=0.4, clip=true, trim=0mm 0mm 0mm 0mm]{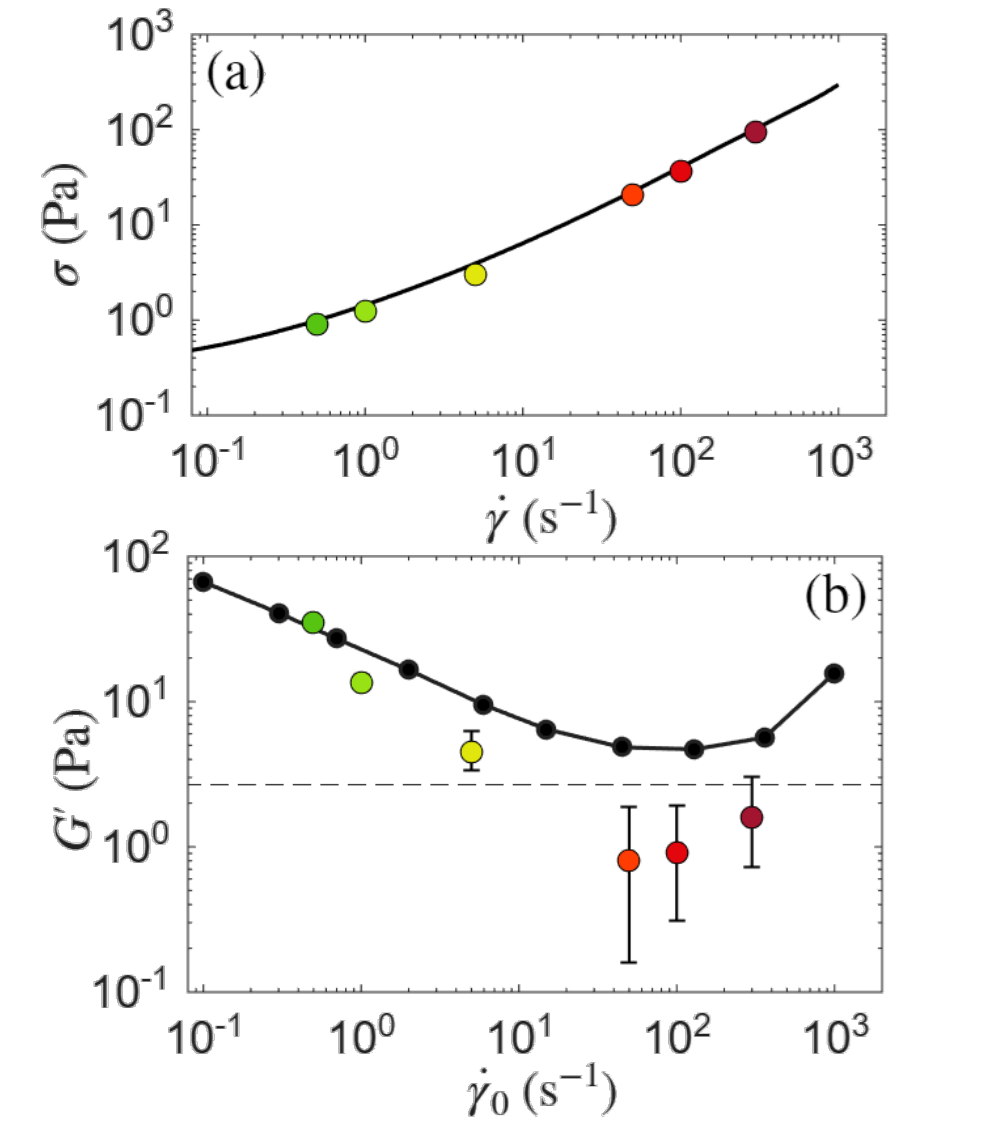}
\caption{Rheological properties of the $1.2~\%$ CB dispersion. (a) Steady-state shear stress $\sigma$ as a function of shear rate $\dot \gamma$. Colored symbols correspond to measurements performed during X-ray tomo-rheoscopy experiments using the $2~\rm mm$ gap 3D-printed geometry. The solid black line shows the flow curve measured by conventional rheometry at $T=25^\circ \rm C$ using a coaxial cylinder geometry in polycarbonate ($1~\rm mm$ gap and $40~\mathrm{mm}$ height). (b) Elastic modulus $G^{\prime}$ of the CB gels measured  $300~\rm s$ after flow cessation as a function of the pre-shear rate $\dot \gamma_0$. Colored and black symbols correspond to X-ray tomo-rheoscopy and standard rheological measurements, respectively. Error bars indicate the standard deviation. The dotted black line indicates the lowest measurable modulus for the X-ray tomo-rheoscopy configuration, due to the low-torque limit of the rheometer. 
}
\label{fig:rheol3per}
\end{figure}

\section{Results}\label{sec:results}

\subsection{Rheology and structure of the 1.2~\% CB gel}
\label{sec:1p2CB}

We first focus on the effect of the pre-shear rate on the rheology of the 1.2~\% CB dispersion. Figure~\ref{fig:rheol3per}(a)-(b) compares the steady-state flow behavior of the dispersion and the elastic properties of the gels formed after flow cessation following exposure to different shear rates. Measurements were performed using both a standard coaxial-cylinders geometry with a gap of $1~\rm mm$ and a height of $40~\mathrm{mm}$ \cite{Bauland:2026}, and the 3D printed X-ray tomo-rheoscopy shear cell described in Section~\ref{sec:Xraytomo} as required for tomographic imaging. 

Figure~\ref{fig:rheol3per}(a) shows that the steady-state shear stress measured with the X-ray tomo-rheoscopy shear cell (colored symbols), after $200~\mathrm{s}$ of shear at a given $\dot{\gamma}_0$, closely follows the flow curve obtained with the standard coaxial-cylinders geometry (black line). This agreement validates the rheological measurements performed in the tomo-rheoscopy configuration despite the markedly different gap and sample height, and confirms the classical shear-thinning behavior of a yield-stress fluid. 

Figure~\ref{fig:rheol3per}(b) reports the elastic modulus $G^{\prime}$ measured $300~\mathrm{s}$ after flow cessation, as a function of the pre-shear rate $\dot \gamma_0$. As previously reported in \cite{Bauland:2026}, the elasticity of the resulting gels shows a non-monotonic dependence on the pre-shear rate. The minimum in $G^{\prime}$ observed at $\dot{\gamma}_0 \approx 40~\mathrm{s}^{-1}$ marks the transition between homogeneous gel structures formed following high shear rates ($\dot{\gamma}_0 > 40~\mathrm{s}^{-1}$), associated with reaction-limited cluster aggregation, and shear-densified percolated networks formed after lower shear rates ($\dot{\gamma}_0 < 40~\mathrm{s}^{-1}$).
Measurements performed with the X-ray tomo-rheoscopy cell qualitatively reproduce this non-monotonic behavior. However, the weaker gels formed following intermediate pre-shear rates ($G^{\prime} \sim 1~\mathrm{Pa}$) lie below the sensitivity limit of the rheometer when using a geometry with a small sample height of $6.0~\rm mm$, as indicated by the horizontal dashed line in Fig.~\ref{fig:rheol3per}(b). Consequently, these measurements are affected by large uncertainties. This limitation highlights the experimental trade-off inherent to tomo-rheoscopy: while tomographic imaging imposes constraints on the sample size, the resulting reduction in measurable torque limits the reliable rheological characterization of weak gels.

\begin{figure*}[t!]
\centering
\includegraphics[scale=0.57, clip=true, trim=20mm 0mm 10mm 0mm]{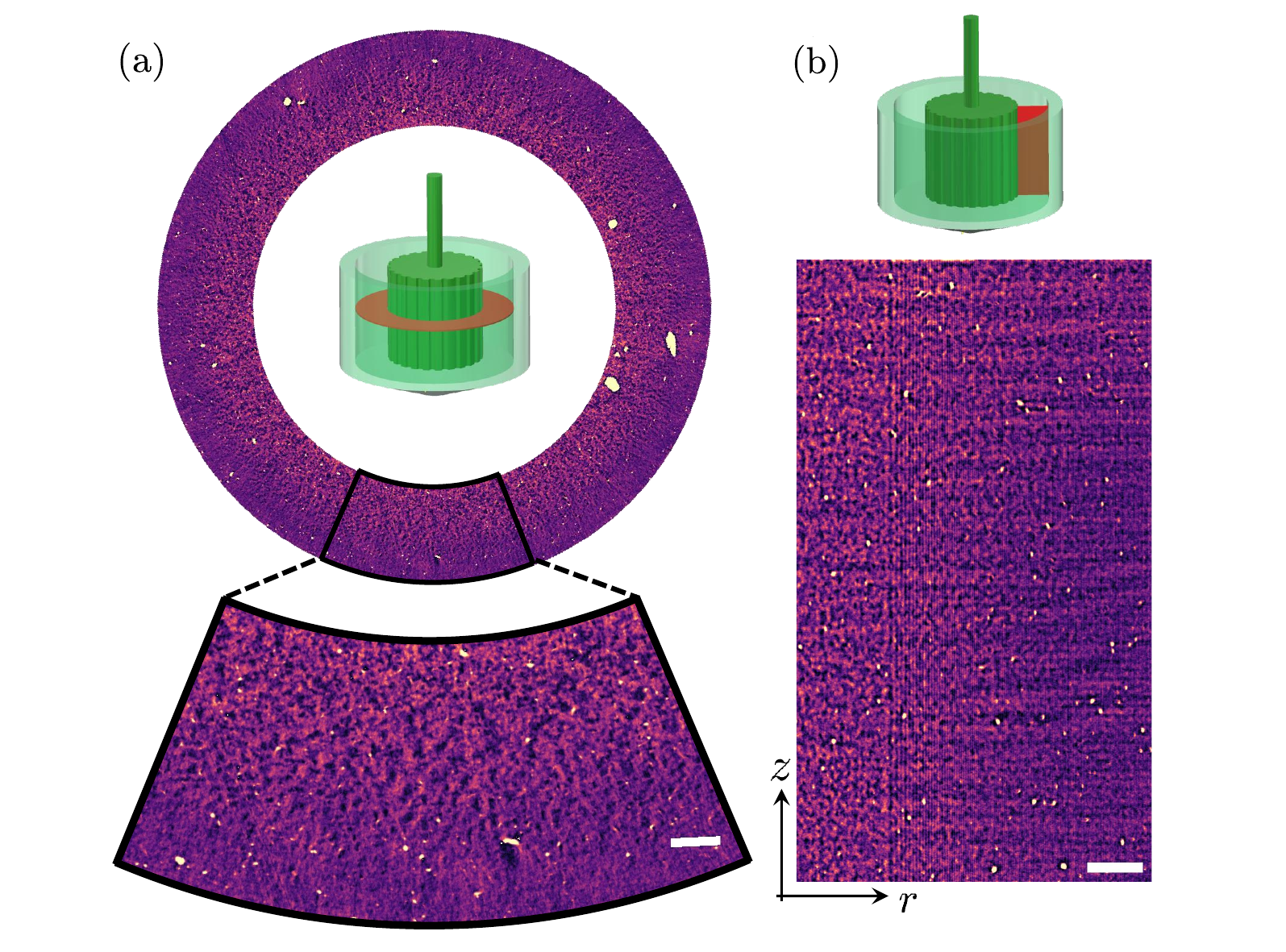}
\caption{Structure of the $1.2~\%$ CB dispersion following a low pre-shear rate ($\dot{\gamma}_0 = 0.1~\mathrm{s}^{-1}$). (a) Horizontal slice taken in the middle of the coaxial cylinders geometry. (b) Orthogonal slice in the radial direction. Scale bar is $300~\mathrm{\mu m}$. For each image, the position in the gap is indicated by a red area in the associated scheme. 
    }
\label{fig:ring3per}
\end{figure*}

To visualize the microstructure of the gel after flow cessation, representative two-dimensional X-ray tomographic slices of the 1.2~\% CB gel microstructure obtained following a pre-shear at the lowest investigated shear rate, $\dot{\gamma}_0 = 0.1~\mathrm{s}^{-1}$ are presented in Figure~\ref{fig:ring3per}. Fig.~\ref{fig:ring3per}(a) presents a horizontal slice of the gap, corresponding to the velocity–velocity gradient plane, while Fig.~\ref{fig:ring3per}(b) shows a vertical slice in the velocity gradient–vorticity plane.  
For visualization purposes, grayscale images are color-mapped such that regions of high-contrast appear bright (yellow–white), while low-contrast regions appear dark (purple–black). In Fig.~\ref{fig:ring3per}(a), a small number of very large particles, with sizes up to several hundred microns, are visible as bright yellow spots, which correspond to impurities originally present in the CB powder \cite{Poozhikunnath:2019}. Beyond these isolated features, the zoomed-in view highlights a textured network composed of particle-rich and particle-poor regions, characteristic of flocculated gel structures. 
This observation provides direct evidence of large-scale structural organization in CB dispersions. 

At this particle volume fraction and low pre-shear rate, the gel structure is clearly heterogeneous across the gap. As visible in Fig.~\ref{fig:ring3per}(a), the microstructure appears significantly coarser in the vicinity of the inner cylinder, while progressively becoming finer and more homogeneous toward the outer cylinder. This radial gradient is further evidenced by the orthogonal slice shown in Fig.~\ref{fig:ring3per}(b), which reveals a modulation of the microstructure along the vorticity direction. This suggests that the gel structure formed during pre-shear exhibits three-dimensional spatial variations, beyond a purely radial heterogeneity.

\begin{figure*}[t!]
\centering
\includegraphics[width=\linewidth]{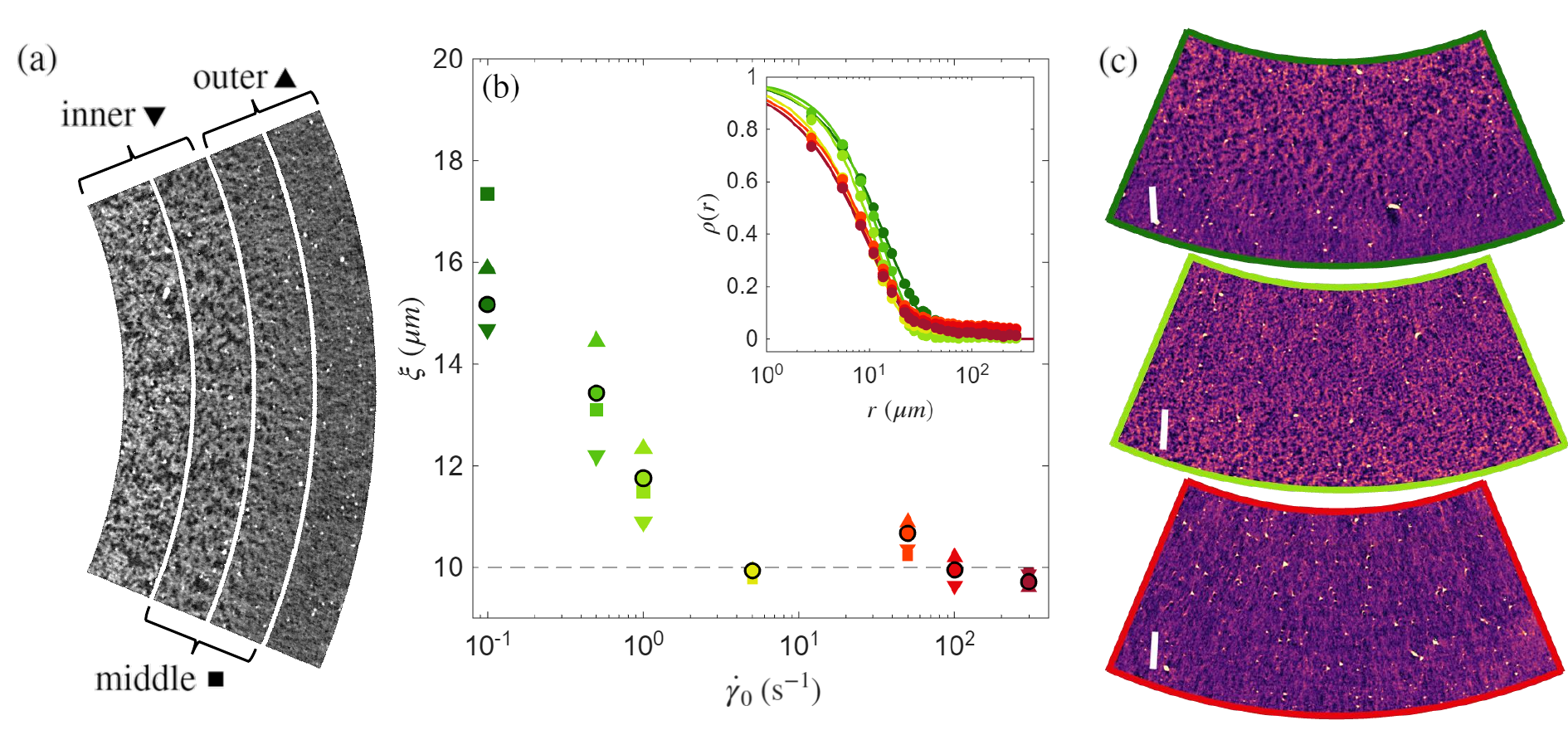}
\caption{(a) Schematic representation of the Couette gap divided into three regions used for image analysis: inner, middle, and outer sections. (b) Correlation length $\xi$ of the $1.2~\%$ CB structure as a function of the pre-shear rate $\dot \gamma_0$. Marker shapes indicate the gap region over which the autocorrelation function was computed: whole gap ($\bigcirc$), inner ($\nabla$), middle ($\square$), and outer ($\triangle$). The inset shows representative autocorrelation functions $\rho(r)$ as a function of radial distance $r$, together with their best exponential fits. (c) Representative reconstructed images of the gel microstructure in the $(r,v)$ plane for the three pre-shear rates. From top to bottom: $\dot{\gamma}_0 = 0.1$, $1$, and $100~\mathrm{s^{-1}}$.
}
\label{fig:corr3per}
\end{figure*}

As suggested by the dependence of the gel viscoelasticity on the pre-shear rate, the microstructure exhibits a clear evolution, as illustrated in Fig.~\ref{fig:corr3per}(c). At low pre-shear rate ($\dot{\gamma}_0 = 0.1~\mathrm{s^{-1}}$), the gel displays a coarse and strongly heterogeneous microstructure dominated by large floc-like domains. Increasing the pre-shear rate to $\dot{\gamma}_0 = 1~\mathrm{s^{-1}}$ leads to a marked refinement of the structure, while at $\dot{\gamma}_0 = 100~\mathrm{s^{-1}}$, the microstructure appears nearly homogeneous at the scale resolved by the X-ray tomography, although residual heterogeneities remain visible.

To quantitatively characterize the length scales associated with these heterogeneous structures, we analyze horizontal slices at fixed height along the vorticity direction, as shown in Fig.~\ref{fig:ring3per}(a), using a spatial autocorrelation approach described in Section~\ref{sec:imageanalysis}. Because the microstructure is not homogeneous across the gap, the latter is divided into three regions --inner, middle, and outer-- as schematically illustrated in Fig.~\ref{fig:corr3per}(a). For each region, as well as for the whole gap, we compute the two-point autocorrelation function $\rho(r)$ of the reconstructed images (see Section~\ref{sec:imageanalysis} for technical details). 
Representative autocorrelation functions are shown in the inset in Fig.~\ref{fig:corr3per}(b). In all cases, $\rho(r)$ is well described by an exponential decay, from which a characteristic correlation length $\xi$ can be extracted. This analysis is repeated for each gap region and for various pre-shear rates $\dot \gamma_0$, and the resulting values of $\xi$ are summarized in Fig.~\ref{fig:corr3per}(b). The correlation length $\xi$ increases markedly as the pre-shear rate decreases, indicating the formation of progressively coarser structures at lower pre-shear rates. 

Importantly, for all pre-shear conditions, the extracted length scales are much larger than the primary particle size: $\xi$ typically exceeds $10~\rm \mu m$, and is about two orders of magnitude larger than the $\sim 200~\rm nm$ diameter of the CB particles. 
For fractal aggregates, the cluster size can also be estimated from ultra-small-angle X-ray scattering (USAXS) through mass conservation arguments even when it exceeds the instrument's measurement window, provided that the particle size, volume fraction, and fractal dimension are known. The value $\xi \approx 10~\mu\rm{m}$ obtained here for gels formed under high pre-shear rates is in good agreement with estimates derived from USAXS measurements \cite{Bauland:2026}, demonstrating the consistency of the structural picture obtained across complementary length-scale-resolved techniques.

Moreover, the comparison between gap regions reveals a systematic structural gradient: for a given pre-shear rate (excluding $\dot{\gamma}_0 = 0.1~\rm{s}^{-1}$), the correlation length is largest in the outer region, intermediate in the middle region, and smallest near the inner cylinder. These trends, reported as distinct symbols in Fig.~\ref{fig:corr3per}(b), provide a quantitative measure of the radial structural heterogeneity induced during the pre-shear step. Notably, the differences in correlation length between the inner, middle, and outer regions progressively diminish as the pre-shear rate increases, indicating that higher pre-shear rates promote a more uniform microstructure across the gap.

\begin{figure}[t!]
\centering
\includegraphics[scale=0.45, clip=true, trim=0mm 0mm 0mm 0mm]{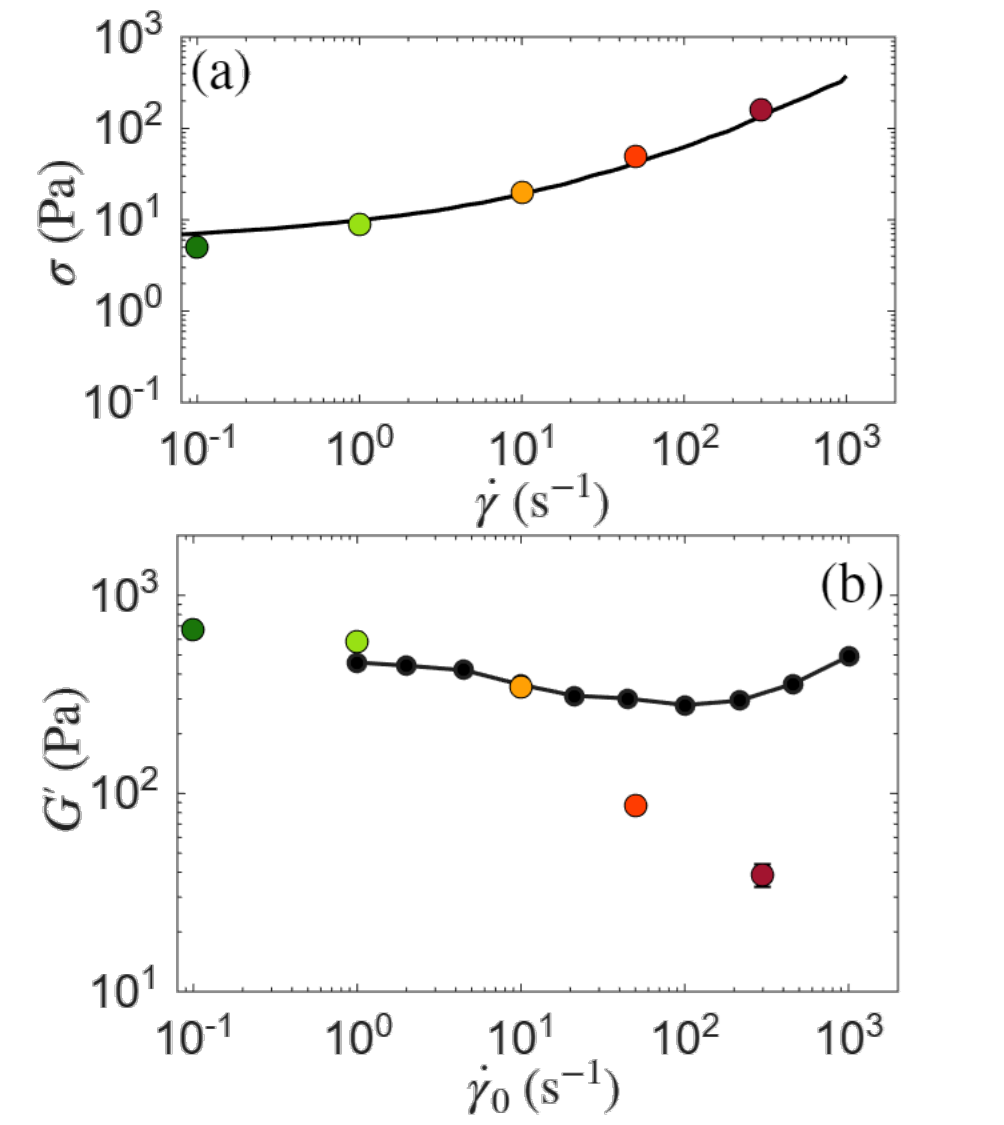}
\caption{Rheology of the $2.4~\%$ CB dispersion. (a) Steady-state shear stress $\sigma$ as a function of shear rate $\dot \gamma$. Colored symbols correspond to measurements performed during X-ray tomo-rheoscopy experiments using the 2~mm gap 3D-printed geometry. The solid black line shows the flow curve measured by conventional rheometry at $T=25^\circ \rm C$ using a coaxial cylinder geometry in polycarbonate ($1~\rm mm$ gap and $40~\mathrm{mm}$ height). (b) Elastic modulus $G'$ of the CB gels measured  $300~\rm s$ after flow cessation as a function of the pre-shear rate. Colored and black symbols correspond to X-ray tomo-rheoscopy and standard rheological measurements, respectively. Error bars indicate the standard deviation.
}
\label{fig:rheol6}
\end{figure}

\subsection{Rheology and structure of the 2.4\% CB gel}

We now turn to the more concentrated CB dispersion at $\phi = 2.4\%$, for which the same combined rheological and imaging protocol was applied. Figure~\ref{fig:rheol6}(a) compares the steady-state flow curves measured using the miniature 3D-printed Couette cell and a standard Couette geometry, in the same fashion as the comparison presented in Figure~\ref{fig:rheol3per}(a) for the $\phi = 1.2$~\% CB gel. Here again, the excellent agreement between the two datasets confirms again the reliability of the rheological measurements performed in the 3D-printed geometry under steady shear.

Figure~\ref{fig:rheol6}(b) shows the elastic modulus $G^{\prime}$ measured $300~\rm s$ after flow cessation as a function of the pre-shear rate $\dot \gamma_0$. In contrast with the 1.2~\% sample, all measured values of $G^{\prime}$ now lie well above the minimum detectable modulus set by the torque resolution of the rheometer. Moreover, the non-monotonic dependence of $G^{\prime}$ on the pre-shear rate is even more pronounced. In particular, when measured in the 3D-printed Couette cell with a $2~\mathrm{mm}$ gap, $G^{\prime}$ decreases down to approximately $40~\mathrm{Pa}$ following a pre-shear at $\dot \gamma_0=300~\rm s^{-1}$, compared to a minimum value of about $300~\rm Pa$ measured in the standard $1~\mathrm{mm}$ gap geometry after a pre-shear at $\dot \gamma_0 \simeq 100~\rm s^{-1}$. This quantitative difference may originate from geometrical factors, including the reduced sample height in the 3D-printed cell ($6.0~\rm mm$ compared to $40~\rm mm$ in the standard geometry), as well as from differences in gap size that may enhance shear-rate gradients. Nevertheless, the key result is that both geometries consistently reveal an increase in the gel elasticity for decreasing pre-shear rates for this more concentrated sample. 

\begin{figure*}[t!]
\centering
\includegraphics[width=\linewidth]{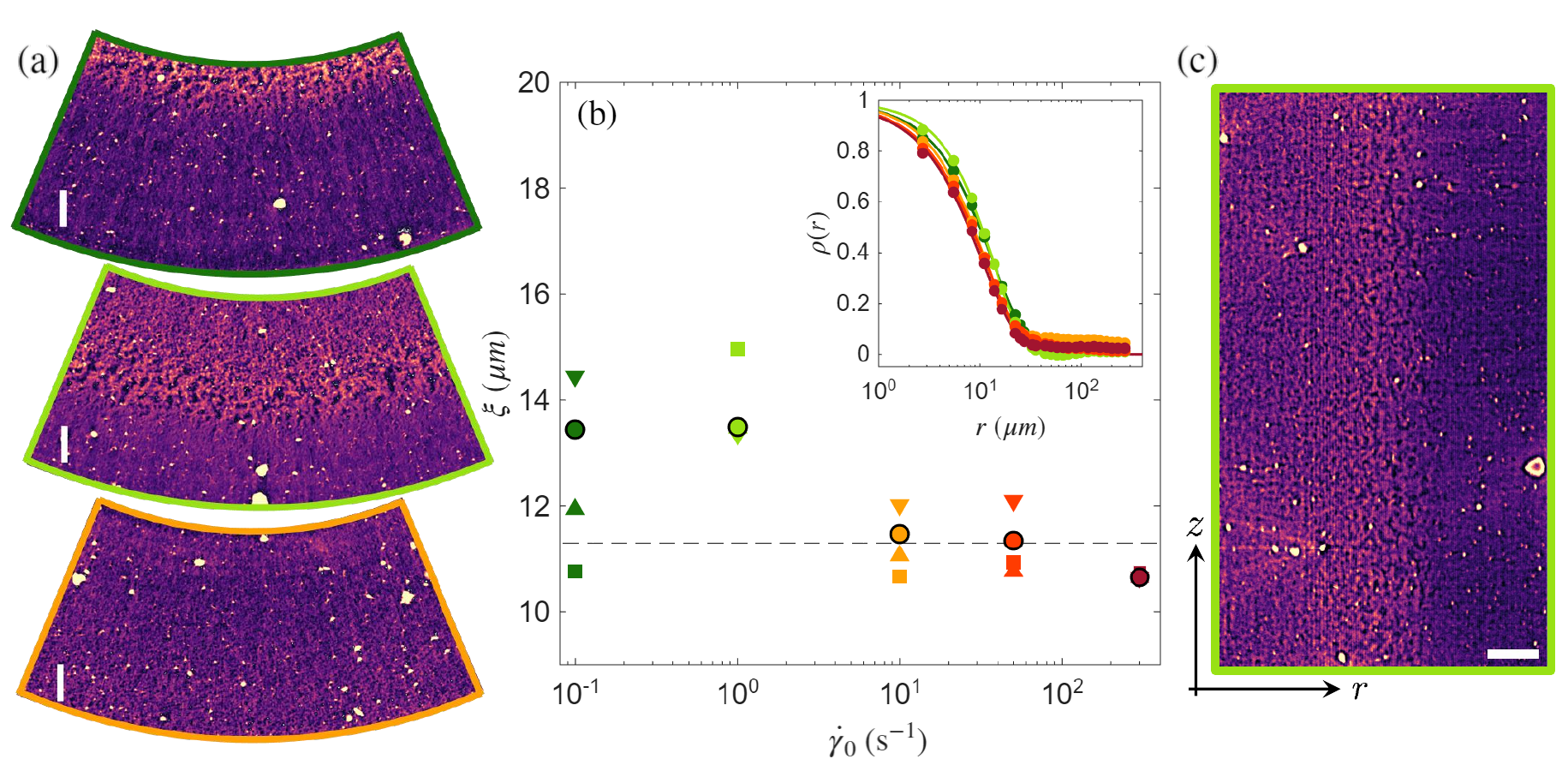}
\caption{(a) Representative reconstructed images of the $2.4~\%$ CB gels in the $(r,v)$ plane following different pre-shear rates. From top to bottom: $\dot \gamma_0 = 0.1$, $1$, and $10~\rm s^{-1}$. The scale bar is $300~\rm \mu m$. (b) Correlation length $\xi$ of the gel microstructure as a function of the pre-shear rate $\dot \gamma_0$. Marker shapes indicate the gap region used for the calculation: whole gap ($\bigcirc$), inner ($\nabla$), middle ($\square$), and outer ($\triangle$). The inset shows representative autocorrelation functions $\rho(r)$ as a function of radial distance $r$, together with their best exponential fits. (c) Reconstructed orthogonal slice in the (z,r) plane along the vorticity direction following a pre-shear rate $\dot \gamma_0 = 1~\rm{s^{-1}}$.
}
\label{fig:corr6per}
\end{figure*}

Representative reconstructed images of the gel microstructure following pre-shear rates of $\dot\gamma_0=0.1$, $1$, and $10~\rm s^{-1}$ are shown in Fig.~\ref{fig:corr6per}(a) [see also Fig.~\ref{figsupp:im}(b) in Appendix~\ref{sec:appendix} for a more complete set of images]. Compared to the $1.2~\%$ sample, spatial heterogeneities are markedly more pronounced. At $\dot \gamma_0=0.1~\rm s^{-1}$, a coarse microstructure is primarily localized in the vicinity of the inner cylinder. At $\dot \gamma_0 =1~\rm s^{-1}$, the heterogeneity becomes even more striking: a coarse, highly structured region develops in the central part of the gap and occupies nearly one third of it, while the regions close to the inner and outer cylinders appear comparatively more homogeneous. This configuration recalls a fault-like structure, qualitatively reminiscent of a shear-localized band embedded between two more uniform regions \cite{BenZion:2003,Bense:2013}. At the highest pre-shear rate investigated, $\dot \gamma_0 =10~\rm s^{-1}$, the microstructure appears more homogeneous overall, although residual coarse domains remain visible at various radial positions.

As for the $1.2~\%$ sample, we quantify these observations by computing spatial autocorrelation functions $\rho(r)$ for images acquired at different pre-shear rates, following the procedure described in Section~\ref{sec:imageanalysis}. The analysis is performed separately in the inner, middle, and outer regions of the gap, as well as over the whole gap. The resulting correlation lengths $\xi$ are reported in Fig.~\ref{fig:corr6per}(b).
For all gap regions, the correlation length increases as the pre-shear rate decreases, confirming that lower pre-shear rates lead to the formation of coarser gel structures. The extracted values of $\xi$ are of the same order of magnitude as those obtained for the $1.2~\%$ sample, typically exceeding $10~\rm \mu m$ and thus remaining much larger than the primary CB particle size. However, the radial distribution of $\xi$ now exhibits a more complex behavior. For most pre-shear rates, the largest correlation length is found in the inner region of the gap, except at $\dot \gamma_0=1~\rm s^{-1}$, where the middle region displays the largest value of $\xi$. At the lowest pre-shear rate, $\dot \gamma_0=0.1~\rm s^{-1}$, pronounced differences are observed between regions, with $\xi$ increasing from the outer to the inner and then to the middle region, reflecting the strong spatial heterogeneity apparent in the images.  
Finally, Fig.~\ref{fig:corr6per}(c) shows a reconstructed slice in the $(z,r)$ plane along the vorticity direction following a pre-shear rate of $\dot \gamma_0=1~\rm s^{-1}$. This vertical cut confirms the presence of a radial structural gradient, while indicating a comparatively more homogeneous organization along the vorticity direction.

\section{Discussion and conclusion}
\label{sec:discussion}

Using X-ray tomo-rheoscopy, we have directly visualized the microstructure of CB gels formed under controlled shear history, unraveling heterogeneities spanning length scales in the tens of microns, far larger than the particle size, and well beyond the range accessible by conventional scattering techniques. 

The analysis of the reconstructed volumes reveals a clear and robust trend valid for both CB concentrations: the structural correlation length $\xi$ decreases with increasing pre-shear rate, up to $\dot \gamma_0 \simeq 5~\rm s^{-1}$, beyond which it reaches a plateau and becomes essentially independent of the pre-shear rate. 
This monotonic evolution of the correlation length contrasts with the non-monotonic dependence of the elastic modulus on the pre-shear rate. While $G'$ initially decreases as the pre-shear rate is increased, it reaches a minimum following a pre-shear rate $\dot \gamma_0 \simeq 100~\rm s^{-1}$ for which the correlation length has already saturated and no longer depends on the imposed flow. At higher pre-shear rates,  $G'$ increases again, despite the absence of any detectable change in the mesoscale correlation probed by X-ray tomography. 
In the high pre-shear rate regime, complementary rheo-USAXS measurements have shown that the structural evolution persists at length scales below the spatial resolution of X-ray tomo-rheoscopy \cite{Bauland:2026}, despite the apparent saturation of the mesoscale correlation length. This work shows that rheological reinforcement at high shear rates is governed by submicrometric structural rearrangements, which remain invisible to tomo-rheoscopic imaging.

Taken together, these observations strongly suggest that a direct correlation between the elastic modulus and the mesoscale structural length scale exists only at low pre-shear-rates. In this limit, decreasing the pre-shear rate leads to the formation of increasingly coarse and heterogeneous structures, characterized by a growing correlation length, which is associated with an increase in the gel elasticity. Such low-shear-rate reinforcement has also been reported in gels of anisotropic particles, such as boehmite platelet suspensions, where it has been linked to the development of shear-induced structural anisotropy \cite{Sudreau:2022,Sudreau:2023}, as well as in gels of charged rigid rods, platelets, and needles \cite{Das:2022,Burger:2024}. In contrast, this regime is notably absent in depletion gels composed of spherical particles \cite{Koumakis:2015}, suggesting that particle shape and/or the nature of interparticle forces play a key role in enabling low-shear-rate reinforcement through mesoscale structural organization. 

In contrast, the increase of $G'$ at high shear rates cannot be accounted for by the mesoscale structural features determined by tomo-rheoscopy. Since $\xi$ remains constant in this regime, the observed rheological reinforcement must originate from structural rearrangements occurring at smaller length scales, below the spatial resolution of about $3~\rm \mu m$ of the present imaging approach. 
These observations are compatible with a scenario in which sufficiently large shear rates fully disrupt the gel microstructure, leading after shear cessation to a more homogeneous gel with enhanced elasticity, as reported for depletion gels \cite{Koumakis:2015}. 

More broadly, our work establishes X-ray tomo-rheoscopy as a powerful tool to bridge rheology and three-dimensional structure in optically opaque soft materials. By granting access to length scales ranging from a few micrometers to hundreds of micrometers under realistic shear conditions, tomo-rheoscopy opens new perspectives to investigate flow heterogeneities, structural memory, and failure mechanisms in a wide class of soft solids, including colloidal gels. Future developments should focus on time-resolved tomo-rheoscopic measurements during the pre-shear step itself, in order to directly capture the dynamical pathways by which large-scale spatial heterogeneities emerge, evolve, or are erased under flow.

\backmatter





\bmhead{Acknowledgements}

The authors thank T.~Gibaud and S.~Manneville for fruitful discussions. The authors acknowledge the Paul Scherrer Institute, Villigen, Switzerland, for providing synchrotron radiation beamtime at the TOMCAT beamline X02DA of the SLS. The Tomo-Rheoscope used in this study was funded by the Swiss National Science Foundation (Grant No. 205311, https://data.snf.ch/grants/grant/205311). The authors also acknowledge the contribution of SFR Sant\'e Lyon-Est (UAR3453 CNRS, US7 Inserm, UCBL) facility: CIQLE (a LyMIC member), especially S. Karpati, for their help with the TEM observations. This research was supported in part by grant NSF PHY-2309135 to the Kavli Institute for Theoretical Physics (KITP).

\section*{Author contribution}

Conceptualization - J.B., T.D.; Methodology - S.G., C.M.S; Formal Analysis - J.B., S.R., C.M.S.; Investigation: J.B., S.G., C.M.S, T.D.; Writing - original draft preparation - J.B., C.M.S, T.D.; Writing - review and editing - J.B., S.R., S.G., C.M.S., M.H., T.D.; Funding acquisition: M.H.; Supervision - M.H., T.D.

%
%


\begin{appendices}
\section{Fraction of outliers in the images}
\label{sec:outliers}

\begin{figure}[b!]
\centering
\includegraphics[width=0.5\linewidth]{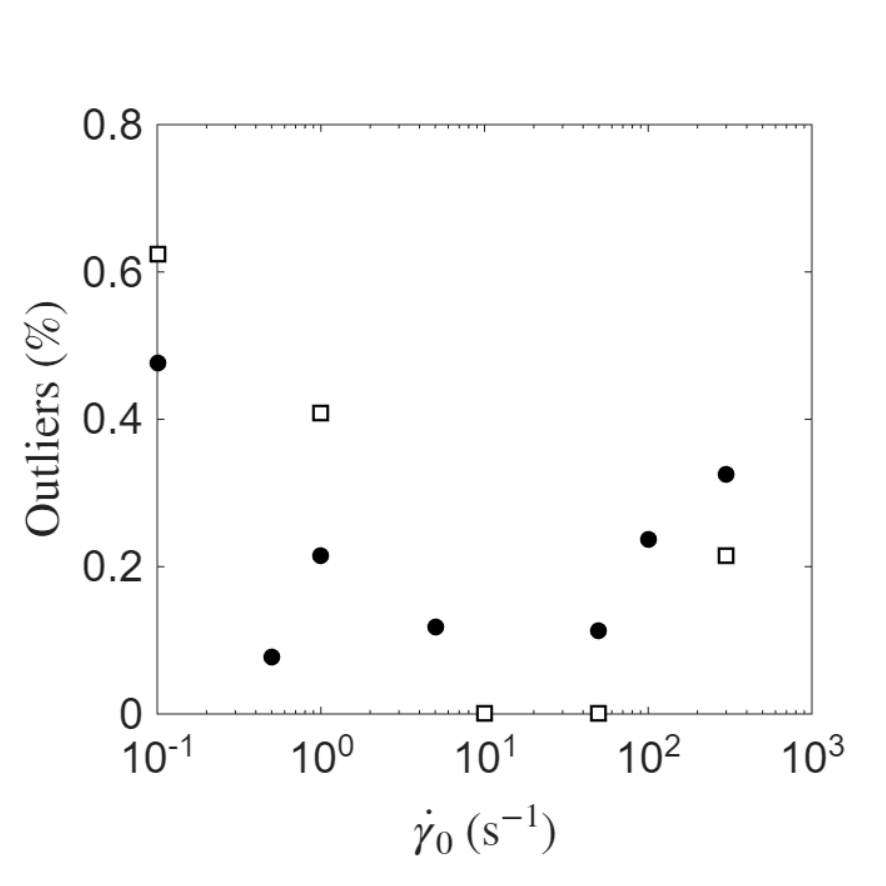}
\caption{Surface fraction of outliers in images corresponding to the $1.2~\%$ (filled circles) and $2.4~\%$ (open squares) CB dispersions at various pre-shear rates.
}
\label{figsupp:outliers}
\end{figure}

Fig.~\ref{figsupp:outliers} displays the surface fraction of high-intensity features associated with impurities in the raw carbon black powder. These metallic impurities (rich in silicon and iron), have been characterized previously by X-ray computed tomography revealing a primary population of 3–5~$\mu$m with larger particles up to 20~$\mu$m, fully consistent with the sizes observed in the present study~\cite{Poozhikunnath:2019}.

For all tested concentrations and pre-shear rates, these impurities represent less than 1~\% of the image area, which is consistent with visual inspection. Interestingly, the fraction of detected outliers exhibits a slight dependence on the pre-shear rate, irrespective of the CB concentration; the fraction of outliers is minimal at intermediate shear rates. Since outliers are identified as pixels with intensities exceeding five times the standard deviation of the radial average intensity, the detection threshold may vary slightly between images.

Given that horizontal slices were taken at the mid-height of the gap, the lower fraction of impurities observed at intermediate shear rates could also be attributed to enhanced sedimentation. This may reflect a competition between the fluidization of the yield stress fluid and redispersion by advection at higher shear rates. Nonetheless, these differences remain minimal and are expected to have a negligible effect on the overall image analysis.

\section{Comparison of reconstructed images obtained at different pre-shear rates}
\label{sec:appendix}

Appendix~\ref{sec:appendix} presents two series of reconstructed images of CB dispersions at two different concentrations, obtained after pre-shear at comparable shear rates. For both concentrations investigated, as discussed in the main text, gels formed at low pre-shear rates exhibit a coarse and heterogeneous microstructure, which becomes increasingly finer and more homogeneous as the pre-shear rate exceeds $10~\rm{s^{-1}}$. In addition, the number of large, bright yellow aggregates increases with CB concentration, supporting their attribution to impurities present in the CB powder.

\begin{figure}[b!]
\centering
\includegraphics[width=0.75\linewidth]{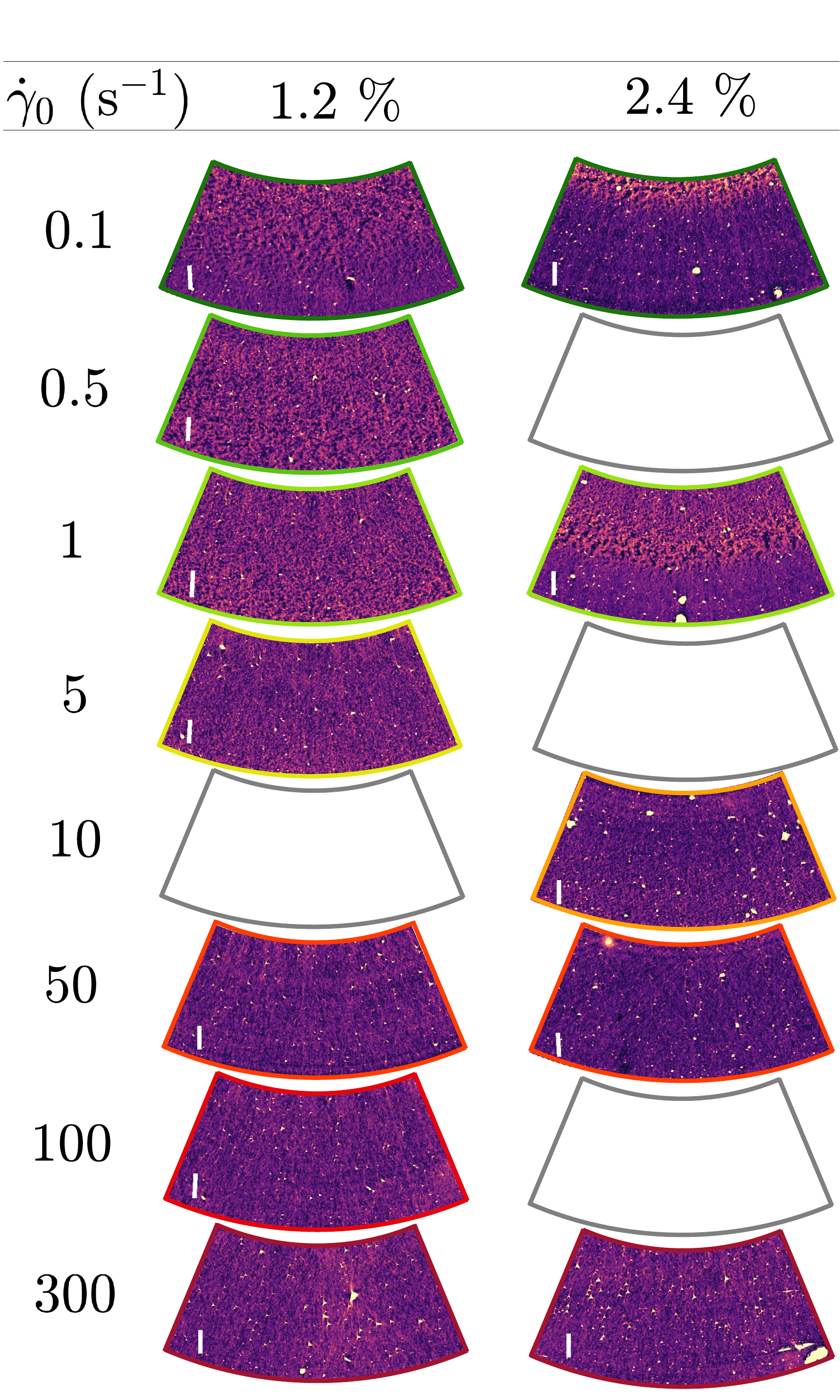}
\caption{Representative reconstructed images of the $1.2~\%$ and $2.4~\%$ CB dispersions following various pre-shear rates. Empty areas indicate combinations of pre-shear rate and particle concentration that were not investigated. 
}
\label{figsupp:im}
\end{figure}

\clearpage




\end{appendices}


\bibliography{biblio}

\end{document}